\documentclass{article}

\usepackage{PRIMEarxiv}

\usepackage[utf8]{inputenc} % allow utf-8 input
\usepackage[T1]{fontenc}    % use 8-bit T1 fonts
\usepackage{hyperref}       % hyperlinks
\usepackage{url}            % simple URL typesetting
\usepackage{booktabs}       % professional-quality tables
\usepackage{amsfonts}       % blackboard math symbols
\usepackage{nicefrac}       % compact symbols for 1/2, etc.
\usepackage{microtype}      % microtypography
\usepackage{lipsum}
\usepackage{fancyhdr}       % header
\usepackage{graphicx}       % graphics
\usepackage{xspace}
\usepackage{amsmath}
\usepackage[normalem]{ulem}
\graphicspath{{media/}}     % organize your images and other figures under media/ folder
\usepackage[numbers, sort&compress, comma, square]{natbib}

\usepackage{comment}
\usepackage{ulem}
\usepackage{subfigure}

\newcommand{\dataset}{\mathcal{E}}
\newcommand{\tr}{\text{train}}
\newcommand{\test}{\text{test}}
\newcommand{\X}{\mathcal{X}\xspace}
\newcommand{\Y}{\mathcal{Y}}
\newcommand{\x}{\mathbf{x}}

\newcommand{\y}{y}
\newcommand{\ie}{\textit{i.e.} }
\newcommand{\eg}{\textit{e.g.} }
\newcommand{\cf}{\textit{c.f.} }
\newcommand{\dom}{\mathcal{D}}
\newcommand{\task}{\mathcal{T}}
\newcommand{\Rep}{\mathcal{R}}
\newcommand{\mcv}{M$^3$CV}
\newcommand{\erpcore}{ERP CORE\space}

\usepackage[table,xcdraw]{xcolor}
\title{Evaluating the structure of cognitive tasks \\
with transfer learning}

\author{Bruno Aristimunha$^{1, 2}$ \quad Raphael Y. de Camargo$^{2}$ \quad   Walter H. Lopez Pinaya$^{3}$\quad \\\textbf{Sylvain Chevallier}$^{1}$  \quad \textbf{Alexandre Gramfort}$^{1}$  \quad \textbf{Cédric Rommel}$^{1, 4}$ \\
\\
$^1$Université Paris-Saclay, Inria, France \\ $^2$Federal University of ABC, Santo Andre, Brazil \\ 
$^3$King’s College London, London, United Kingdom \\ 
$^4$Valeo.ai, Paris, France\\
}

\begin{document}
\maketitle

\begin{abstract}

Electroencephalography (EEG) decoding is a challenging task due to the limited availability of labelled data.
While transfer learning is a promising technique to address this challenge, it assumes that transferable data domains and task are known, which is not the case in this setting.
This study investigates the transferability of deep learning representations between different EEG decoding tasks. We conduct extensive experiments using state-of-the-art decoding models on two recently released EEG datasets, \erpcore and \mcv, containing over 140 subjects and 11 distinct cognitive tasks. We measure the transferability of learned representations by pre-training deep neural networks on one task and assessing their ability to decode subsequent tasks.
Our experiments demonstrate that, even with linear probing transfer,
significant improvements in decoding performance can be obtained, with gains of up to 28\% compare with the pure supervised approach.
Additionally, we discover evidence that certain decoding paradigms elicit specific and narrow brain activities, while others benefit from pre-training on a broad range of representations. By revealing which tasks transfer well and demonstrating the benefits of transfer learning for EEG decoding, our findings have practical implications for mitigating data scarcity in this setting. 
The transfer maps generated also provide insights into the hierarchical relations between cognitive tasks, hence enhancing our understanding of how these tasks are connected from a neuroscientific standpoint. 

\end{abstract}

% keywords can be removed
\keywords{Transfer Learning \and Deep Learning \and Electroencephalography \and Event-Related Potential  \and Brain-Computer Interfaces}

\section{Introduction}
%
%%%%%%%%%%%%%%%%%%%%%%%%%%%%%%%%%%%%%%%%%%%%%%%%%%%%%%%%%%%%%%%%%%%%%%%%%%%%%%%%%%%%%%%%%%%%%%%%%%%%%%%%%%%%%%%%%%%%%%%%%

While brain encoding consists in predicting brain activations given a certain stimulus, brain decoding tackles the inverse problem: translating recorded neural activity into its originating stimulus or behavior \cite{King:2020}.
This stimulus or behavior can be a visual or auditory element presented, the subject internal mental state (e.g. sleep stage), or the cognitive or motor task being performed during the experiment.
Brain decoding has several important applications, such as
the diagnosis of
neurological disorders \cite{roy2019chrononet, ruffini2019deep, Gemein2020}, the detection of seizures \cite{acharya2013automated, acharya2018deep},
automatic processing of polysomnographic recordings \cite{aboalayon2016sleep} and
brain-computer interfaces
\cite{lotte2018review, van2012brain, lecuyer2008brain}, among others.
Electroencephalography (EEG) is a common and affordable way to record the neural activity in this context \cite{casson2019wearable}. It has the benefit of being non-invasive, having very high time resolution compared to functional magnetic resonance imaging (fMRI) and not requiring a complex and costly infrastructure such as magnetoencephalography (MEG).
In recent years, there has been an increasing interest in using deep learning (DL) models for EEG decoding \cite{roy2019deep}.
As an
example, DL has been shown to be the gold standard when it comes to automatic sleep stage classification \cite{chambon2018deep, perslev2021u, XSleepNet:2022} and has also demonstrated impressive performances in brain-computer interfaces \cite{lawhern2018eegnet, santamaria2020eeg, braindecode:2017, wilson2023deep}.

Unfortunately, DL models are notorious for being data-hungry to extract generalizable discriminative representations.   
This characteristic can be a problem when it comes to EEG decoding since
the acquisition of labelled EEG data remains a constraint, resulting in
datasets of limited size. 
Indeed, EEG annotation often requires a specialist to run experiments
or visually inspect recorded signals for specific patterns.
In addition,
EEG signals have a very low signal-to-noise ratio,
% small size dataset and scarce data,
especially when we compare to other fields of application which were DL thrived, such as image, speech, and text.
This characteristic makes EEG decoding even more challenging for DL methods in the context of small datasets.

A common technique in DL for dealing with scarce data scenarios is transfer learning (TL),
which consists in applying what you have learned in one context to another \cite{wu2022}. 
In other words, it can be used to leverage a large dataset
to improve the performance in a related smaller dataset, making it a promising technique to alleviate the lack of EEG decoding
data.
However, TL assumes the knowledge of transferable data domains and tasks, which is not fully understood when it comes to brain data.
Indeed, even beyond EEG decoding,
understanding hierarchical relations between cognitive tasks remains a core question in neuroscience.

\begin{figure}[t]
  \centering
  \subfigure[ShallowNet]{
    \includegraphics[width=0.3\linewidth]{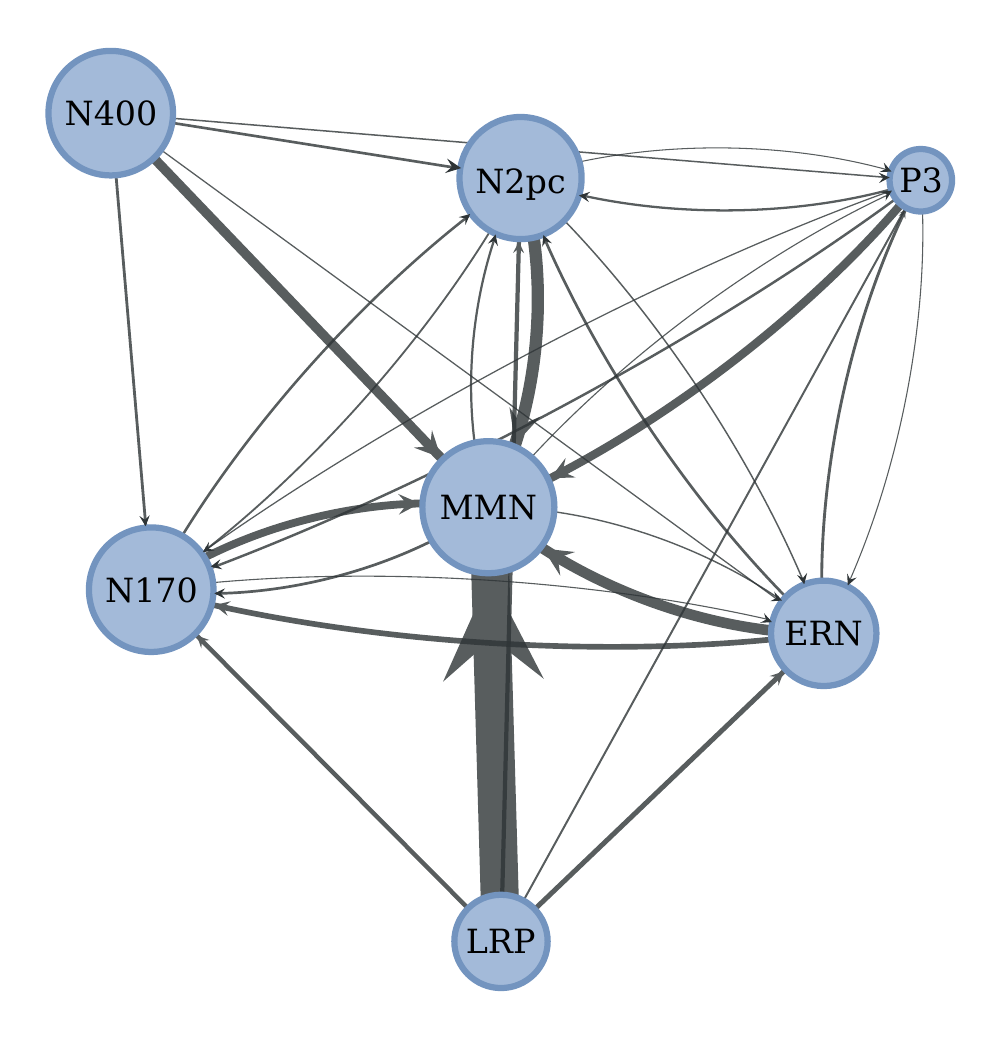}
  }
  \subfigure[EEGNet]{
    \includegraphics[width=0.3\linewidth]{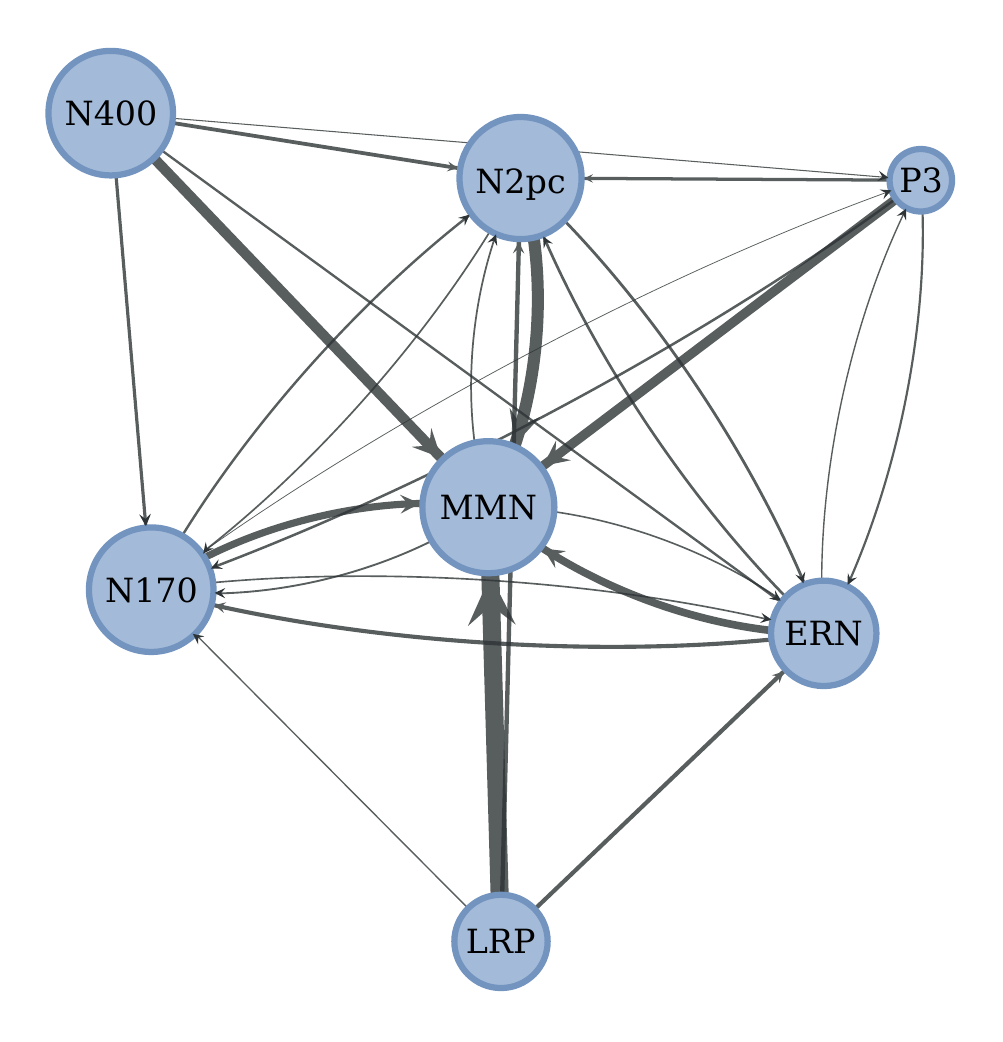}
  }
  \subfigure[EEGInception]{
    \includegraphics[width=0.3\linewidth]{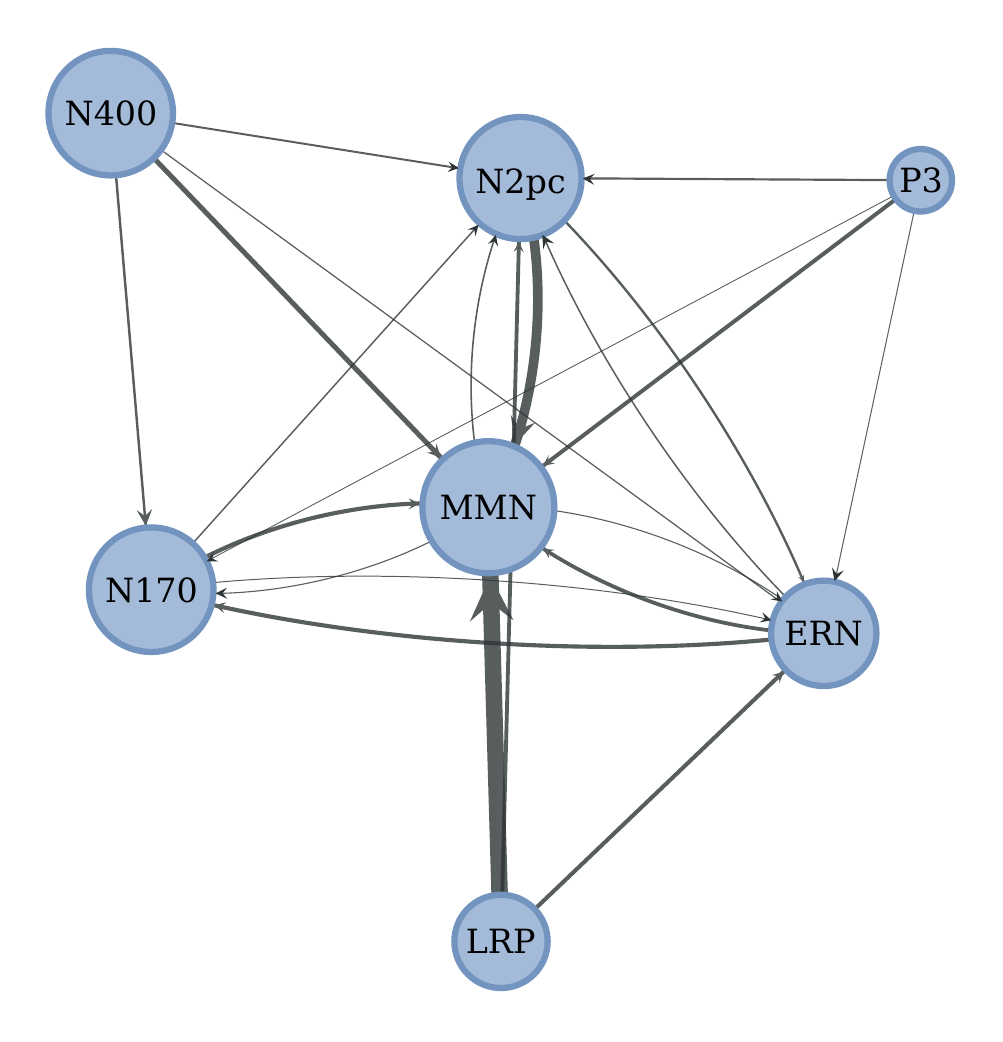}
  }
  \caption{
  Revealing cognitive connections through transfer learning.
  This graph depicts the transferability of representations used for EEG decoding, capturing the intricate interplay between cognitive tasks.
  Each node corresponds to a distinct paradigm within the \erpcore dataset. Arrow width represents the average transfer performance when using the representations learned from a source task to decode a target task.
  MMN is a great target task for transfer learning, benefiting from representations learned from all source tasks.
  On the contrary, LRP is mostly a source task: its representations are useful for a wide range of other paradigms.
  }
  \label{fig:network_graph_erpcore}
\end{figure}

Inspired by the Taskonomy study \cite{zamir2018taskonomy} from the computer vision field, this work investigates the relations between cognitive tasks in an EEG decoding setting.
Specifically, we measure the transferability 
of
representations
learned by DL models
between cognitive tasks.
This transferability is measured by
pre-training DL models to decode the EEG signals of subjects carrying certain tasks and assessing how well a classifier can reuse the learned representations to decode a subsequent task.
We carry extensive experiments with three state-of-the-art decoding models \cite{braindecode:2017, lawhern2018eegnet, santamaria2020eeg} trained and evaluated on two recently released EEG datasets, \erpcore~\citep{kappenman2021erp} and \mcv~\citep{Huang2022}, containing in total over 140 subjects with 11 distinct decoding modalities. 
This enables us to create
transfer maps capturing the relationships between pairs of cognitive tasks, as presented in \autoref{fig:network_graph_erpcore}.
From an EEG processing perspective, our maps can be used to leverage related datasets for alleviating EEG data scarcity with transfer learning.
We show that even with a linear probing transfer method, we are able to boost by up to $28\%$ the performance of some tasks.
From a neuroscientific standpoint,
our results broaden our understanding of the connections between cognitive tasks.
We discover evidence that some decoding paradigms elicit very specific and narrow brain activities, since no other paradigm transfer well into them. On the other hand, the decoding of some cognitive tasks benefits from pre-training on all other paradigms, thus demonstrating that they rely on a broad range of representations.

We organize the paper as follows.
First, \autoref{sec:related} introduces the relevant related work in transfer learning, EEG decoding and cognitive tasks structure investigation. In \autoref{sec:meth}, we formalize
transfer learning for EEG decoding.
Then, \autoref{sec:exp} outlines our experimental setting,
including a description of
the two EEG datasets
and their respective decoding modalities,
as well as our training and data processing protocols.
Subsequently,
we present the main results of our experiments
in \autoref{sec:resu},
highlighting the significant improvements in decoding performance achieved through transfer learning. Finally, in \autoref{sec:discussion}, we discuss the broader implications of our findings and offer insights into the hierarchical and asymmetric relations between cognitive tasks.

\section{Related Works}\label{sec:related}

\subsection{Transfer Learning and Taskonomy}

Transfer learning refers to the technique of leveraging knowledge acquired from one source domain and task
to enhance the performance
in another target domain and task
\cite{pan2010survey, Zhuang2020surveytrans}. 
Domain adaptation, domain generalization and self-supervised learning are hence considered sub-fields of TL under this broad definition \cite{Zhuang2020surveytrans, domainadapt2020, deMathelin2023}.
The most common TL approach consists in fine-tuning on a target dataset a model that was pre-trained on a source dataset.
In this context, the main factor determining the success of TL is the relationship between source and target domains and tasks.
TL might indeed  hinder learning performance in some situations, such as when the source and target are unrelated, which is known as negative transfer~\cite{rosenstein2005transfer, wang2019characterizing}.

Numerous studies have examined the relationship between tasks for TL purposes \cite{zamir2018taskonomy, pmlr-v119-standley20a, Arghya:2019, Dwivedi:2019, Achille:2019, wang}.
One of the most influential works in this area is Taskonomy~\cite{zamir2018taskonomy}, which investigates these relationships in the context of computer vision tasks.
Although we take great inspiration from it, 
our works differ
primarily in the definition of what a "task" is.
While Taskonomy tries to uncover the relation between \emph{visual learning tasks} (\ie \emph{given an image, what is the visual label?}), we focus in relating \emph{cognitive tasks} (\ie \emph{given the measured brain activity, what is the subject doing?}).
In this sense, while Taskonomy works with the same input images, analyzing transfer to different output distributions $P_S(Y) \to P_T(Y)$, our setting is more challenging as we need to transfer between joint distributions of input EEG signals and output decoded stimuli 
$P_S(X,Y) \to P_T(X,Y)$.
This distinction is made clearer in \autoref{sec:meth}.

\subsection{Transfer learning with brain data}

Most EEG decoding works involving transfer learning focus on cross-subject evaluation within the same decoding task \cite{LI2023195, lotte2010learning, devlaminck2011multisubject, jayaram2016transfer, Jin2022, Wei2023, gao2023double, Gao:2023, Qingshan:2022, wu2022, kalunga2018transfer, Qingshan:2023, Apicella2022OnTE}. %
Given the extremely high inter-subject variability of EEG signals, this is a crucial question in the design of real-world systems able to generalize from one subject to another.
While we study
the transfer between different tasks, previous studies consider the same fixed task in different data domains corresponding to each subject or session. They enter in the more specific sub-category of domain adaptation and generalization.

Most related to our work, \cite{Antonello2021, oota2022, wang, qu2023} measure the transferability between cognitive tasks, drawing inspiration from the Taskonomy framework~\cite{zamir2018taskonomy}.
However, these works are all based on fMRI data and most use \emph{encoding} models for very specific types of visual or language stimuli.
In contrast, our work uses EEG \emph{decoding} models to compare a broader range of stimuli in different modalities.
By working with decoding instead of encoding models, our results are not only useful to understand the relation between cognitive tasks, but also to improve the performance of automatic EEG processing systems in real-world scenarios where data is scarce.

Similar to our work, \citet{qu2023} also uses decoding models to create a transferability map between a large set of cognitive tasks with fMRI data.
By working with EEG data rather than fMRI, our analysis can better detect signal-level patterns at the expense of a smaller spatial precision, making it complementary to this previous study.
Moreover, while all experiments in \cite{qu2023} were conducted on a single dataset with a small fully connected network architecture, our experimental analysis encompasses consistent results on two very distinct datasets
% with 140 subjects in total
using three different state-of-the-art deep neural network architectures.

\subsection{Uncovering cognitive tasks relations with EEG}

Beyond the framework of transfer learning, 
researchers in the EEG decoding community have
been interested in identifying the latent structure between cognitive tasks by other means.
% using DL.
For example, \citet{cedric2022benchmark} analyzed what data augmentations work best for decoding different cognitive tasks.
They showed that some augmentations work well for many different tasks, demonstrating that they share some common invariances, while others only improve the decoding performance in some specific cases.
Although some structure arises as a by-product of this study, it focuses on task invariances.
Our work differs substantially from \cite{cedric2022benchmark} since our analysis directly connects one cognitive task to another based on transferred predictive performance instead of invariances.

As another example,
studies conducted by \citet{banville2021uncovering, banville2019} have demonstrated that self-supervised representations can encode clinically-relevant structures from EEG data, such as the sleep stages, pathology, age, apnea and gender information, without any access to such labels.
A similar outcome was observed in \citep{azabouusing:2021, mohsenvand:2020}. 
While these methods are capable of learning representations useful for different downstream tasks, these works do not study
transfer
to different domains and datasets in practice.
Our work focuses not only on discovering the structure between predefined supervised tasks but also quantitatively assesses the transferred performances into new data domains and tasks.

%%%%%%%%%%%%%%%%%%%%%%%
\section{Method}\label{sec:meth}

% In this section, we will explore the details of our decoding step, the evaluation of transfer learning, and the assessment of transferability using linear probing. To provide a visual representation of our methodology, we present a comprehensive overview in \autoref{fig:overview}.

\begin{figure}[!ht]
    \centering
   \includegraphics[width=\linewidth]{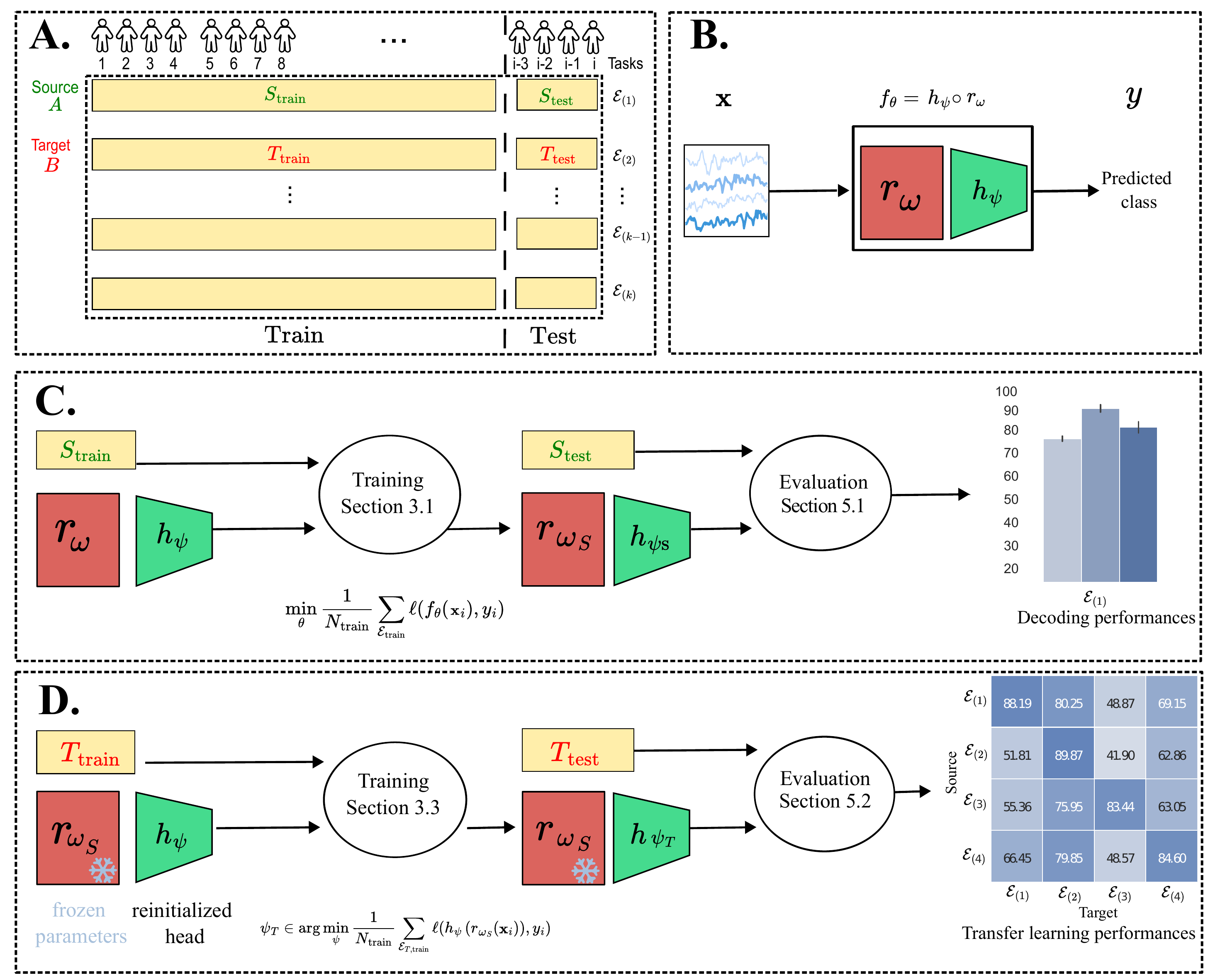}
    \caption{\textbf{A.}~Data splitting and alignment. Source and target tasks correspond to different ERP and BCI paradigms; \textbf{B.}~EEG decoding models as a representer network and a classification head; \textbf{C.}~Standard EEG decoding training and evaluation; \textbf{D.}~Transfer with linear probing. Only the classification head $h_{\psi}$ is re-trained, while the representer network $r_{w_S}$ trained on the source task is kept intact.}
    \label{fig:overview}
\end{figure}

\subsection{EEG Decoding}

% Decoding = classification
From a machine learning perspective, EEG decoding is defined as a classification problem
% Notation dataset
using a dataset of $N$ pairs of trial recordings and decoding labels $\dataset=\{(\x_i, \y_i)\}_{i=1}^N$.
Each trial recording $\x \in \X=\mathbb{R}^{m\times c}$ is modelled as a matrix with $m$ rows and $c$ columns, where $m$ is the number of time-steps and $c$ is the number of electrodes.
The decoding labels $\y \in \Y$ correspond to the class of stimulus to predict, where $\Y$ is a discrete set containing 2 or 3 classes in the decoding scenarios studied in this paper.
We assume the dataset is randomly split into a training set $\dataset_\tr$ of size $N_\tr$ and test set $\dataset_\test$ of size $N_\test$.
The decoding problem amounts to using the training set $\dataset_\tr$ to learn a model $f_\theta: \X \to \Y$ with parameters $\theta$ mapping each trial $\x$ to the associated label $\y$. 
This is achieved by optimizing the model parameters to minimize an average loss function $\ell$ across the training dataset:
\begin{equation}\label{eq:learning-pb}
    \min_\theta \frac{1}{N_\tr} \sum_{\dataset_\tr} \ell( f_\theta (\x_i), \y_i) \enspace .
\end{equation}
In all our experiments, the loss function $\ell$ used is the balanced cross-entropy.
The generalization performance of the trained model is assessed on the test set $\dataset_\test$ (\cf \autoref{fig:overview} A, B, C).

\subsection{Transfer learning}

More generally, we assume that our dataset corresponds to observations of a joint random variable $(X, Y)$ valued on a space $\X \times \Y$, where some decision function $f$ exists such that $Y=f(X)$.
Hence, problem \eqref{eq:learning-pb} amounts to trying to approximate the decision function $f$ by optimizing the model $f_\theta$ to fit the training data sampled from $P(X, Y)$.

Following \cite{Zhuang2020surveytrans}, a \emph{domain} is defined by a feature space and inputs distribution $\dom = \{\X, P(X)\}$.
Likewise, a task consists of an output space and a decision function $\task = \{\Y, f\}$.
Transfer learning aims to use the knowledge extracted from a source domain $\dom_S$ and task $\task_S$ 
to improve the performance of a model $f_\theta$ on another target domain $\dom_T$ and task $\task_T$.
Both domains and tasks are materialized by source and target datasets
${\dataset_S=\{ (x_i, y_i) \sim P_S(X, Y); Y=f_S(X)\}}$ and
${\dataset_T=\{ (x_i, y_i) \sim P_T(X, Y); Y=f_T(X)\}}$.

In our context of EEG decoding, the source and target datasets correspond to different ERP and BCI cognitive tasks performed by the same cohort of subjects in comparable experimental settings (\cf \autoref{sec:exp}).
Hence, while source and target domains share the same feature space $\X=\mathbb{R}^{m \times c}$, they differ in terms of marginal distributions of input trial recordings $P_S(X) \neq P_T(X)$.
Regarding learning tasks, source and target datasets differ in terms of decision functions $f_S \neq f_T$ and output spaces $\Y_S \neq \Y_T$.
As opposed to our setting, Taskonomy~\cite{zamir2018taskonomy} works with a common source and target domains $(\X_S, P_S(X))=(\X_T, P_T(X))$.

\subsection{Transferability through linear probing}
\label{sec:linear-probing}

We evaluate the transferability between source and target datasets through \emph{linear probing}~\cite{alain2016understanding, chen2021empirical}, as described below.
We assume that $f_\theta$ is a neural network made of two parts: a representer model $r_\omega: \X \to \Rep$, with parameters $\omega$, and a classifier head $h_\psi: \Rep \to \Y$, with parameters $\psi$
\begin{equation}
f_\theta (x) = h_\psi \left( r_\omega(\x) \right).
\end{equation}
The parameters $\theta$ of the model are hence the concatenation of $\omega$ and $\psi$.
While the representer $r_\omega$ is responsible for learning useful representations for the learning task, the classifier head $h_\omega$ is just the last linear layer of the network used to deliver the classification decision.
When assessing the transferability between a source and target datasets (\cf \autoref{fig:overview}):
\begin{enumerate}
\item We first train the whole model $f_{\theta_S}=h_{\psi_S} \circ r_{\omega_S}$ on the training split of the source dataset $\dataset_{S, \tr}$ by solving equation \eqref{eq:learning-pb} ;
\item Then we freeze the representer parameters $\omega_S$ and retrain the classifier head $h_{\psi_T}$ from scratch on the training split of the target dataset $\dataset_{T,\tr}$:
\begin{equation}
\psi_T \in \arg\min_\psi \frac{1}{N_\tr} \sum_{\dataset_{T,\tr}} \ell( h_\psi \left( r_{\omega_S} (\x_i) \right), \y_i) \,;
\end{equation}
\item Finally, we evaluate the network obtained $h_{\psi_T} \circ r_{\omega_S}$ on the test split of the target dataset $\dataset_{T, \test}$ and use this metric to assess the transferability between $S$ and $T$.
\end{enumerate} 

We evaluate the transferability through linear probing since it assesses whether the representations learned from the source dataset allow us to classify the target data.
In contrast, fine-tuning (\ie pushing the training of the whole model $h$ and $r$ further with target data) would modify the learned source representations, which would complicate the analysis \cite{kumar2022fine}.

\section{Experiments}
\label{sec:exp}

\subsection{\erpcore and \mcv datasets}\label{sec:datamultisub}

During EEG decoding experiments, subjects perform cognitive tasks with stimuli that evoke specific brain signatures. 
When the nature of the stimuli is similar, they can be categorized into the same paradigm.
EEG decoding studies have been interested in a very large and diverse number of paradigms, which can be categorized as exogenous (where an external stimulus is used \eg event related potential) or endogenous (where
stimuli are induced by a predetermined mental task or behavior, \eg motor imagery) \cite{WOLPAW2002767, lotte2018review}. Only a small number of EEG datasets contain recordings in a diverse set of paradigms with the same subjects and configurations.
Most existing datasets include a limited number of subjects \cite{Allison2010, EmotionMeter:2019} or a limited number of cognitive tasks \cite{bed2021, cho2017, PhysioBank:2000}.

In our study, we use two of the few EEG datasets that
explore a large diversity of paradigms with a
single large cohort of subjects in comparable experimental settings.
The first dataset, \erpcore~\cite{kappenman2021erp}, comprises $40$ subjects ($25$ females and $15$ males between $18$ and $30$ years old). It focuses on exogenous paradigms, featuring seven isolated tasks eliciting specific event-related potential components (ERP), namely Active Visual Oddball (P3b), Word Pair Judgement (N400), Face Perception (N170), Passive Auditory Oddball (MMN), Flankers (LRP and ERN) and Simple Visual Search (N2pc). 

\begin{figure}[!ht]
    \centering
    \includegraphics[width=\linewidth]{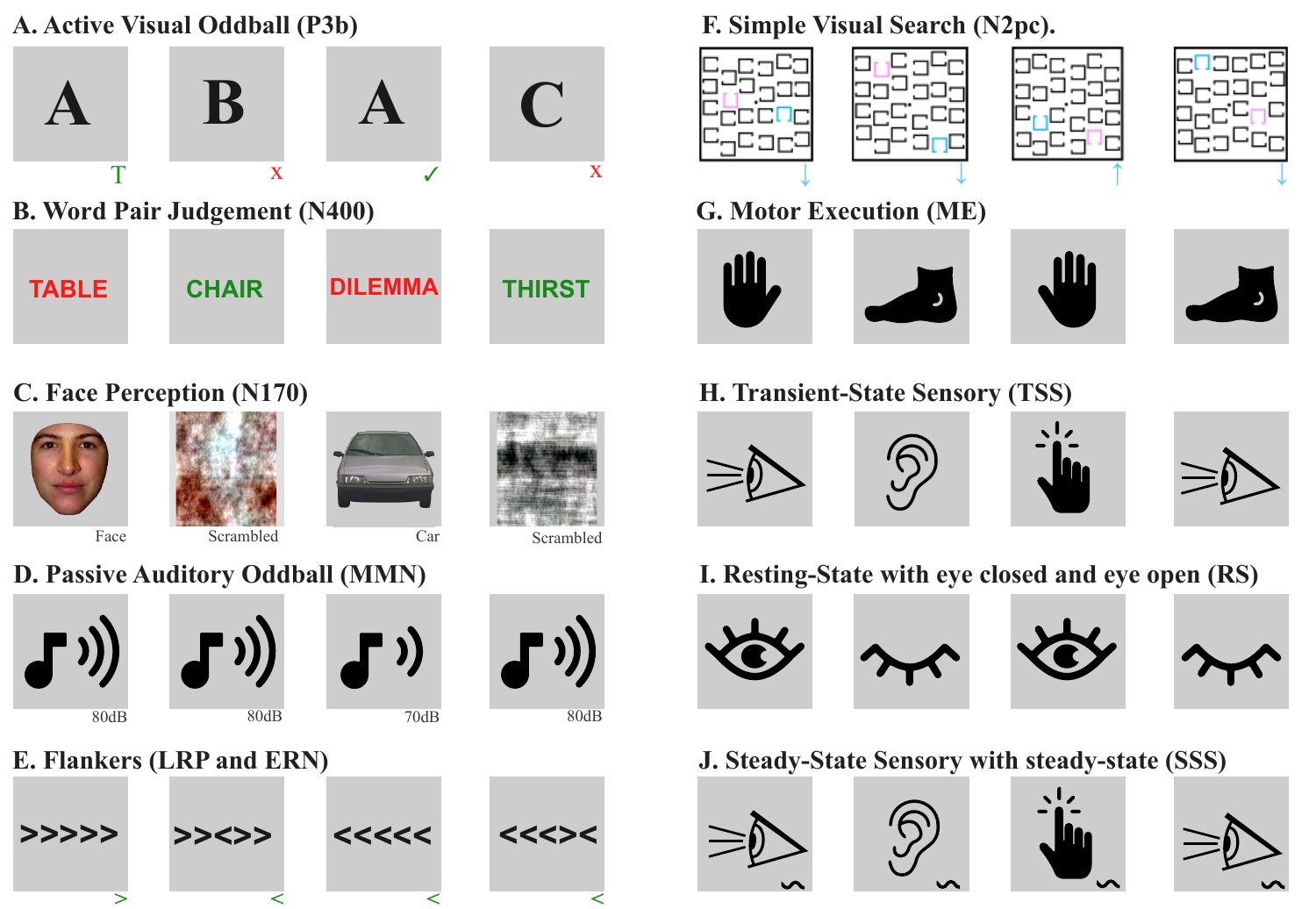}
    \caption{
    Illustration of experiments recorded in \erpcore (\textbf{A} to \textbf{F}) and \mcv datasets (\textbf{G} to \textbf{J}).
    Labels under the stimuli correspond to what subjects are supposed to answer during the trials.
    They do not necessarily correspond to decoding labels.
    \textbf{A.}~In the active visual oddball task P3b, participants viewed random letters and responded to whether each stimulus was matched to the target (T) or not. One of the letters was designated as the target for each block of trials, with a probability of $0.2$.
    \textbf{B.}~The word pair judgment task N400, involved participants viewing a red prime word followed by a green target word and indicating whether the target word was semantically related or unrelated to the prime word. 
    \textbf{C.}~During the face perception task N170, participants viewed images of faces, cars, scrambled faces, or scrambled cars and indicated whether the stimulus was an "object" (face or car) or a "texture" (scrambled face or scrambled car). 
    \textbf{D.}~The passive auditory oddball task MMN involved participants watching a silent video while speakers played standard and deviant tones. Deviant tones had a probability of $0.2$.
    \textbf{E.}~In the flanker's task, participants indicated the direction of the central arrowhead, which was surrounded by arrowheads pointing in the same (congruent trials) or opposite (incongruent trials) directions.
    This task elicited the lateralized readiness potential (LRP) and the error-related negativity (ERN).
    \textbf{F.}~In the simple visual search task N2pc,
    either pink or blue was designated the target colour at the beginning of each trial block.
    Participants had to indicate whether the gap was on the top or bottom of the coloured square.
    \textbf{G.}~During the Motor Execution task (ME), participants responded to a visual cue by \emph{executing} movements with their left hand (LH), right hand (RH), or right foot (FT).
    \textbf{H.}~Transient-State Sensory (TSS) potentials were elicited by visual, auditory, and somatosensory stimuli.
    All three categories of stimuli were arranged in random order and lasted 50 milliseconds each.
    \textbf{I.}~In the Resting-State task (RS), participants completed two runs with eyes closed (EC) and two with eyes open (EO), each lasting one minute.
    \textbf{J.}~The Steady-State Sensory (SSS) task involved separate trains of visual, auditory, and somatosensory stimuli, with different stimulation frequencies and recording times for each run.}
    \label{fig:multidataset}
\end{figure}

The second dataset, \mcv~\citep{Huang2022}, is a large multi-task, multi-session, multi-subject investigation of EEG commonality and variability.
It includes $106$ subjects who performed specific tasks in six different paradigms. 
In this work, we only focus on trials for which the subject and task labels were available, 
namely: Motor Execution (ME), Transient-State Sensory (TSS), Resting-State (RS) and Steady-State Sensory (SSS).
This reduces our dataset to $95$ subjects ($22$ females and $73$ males between $19$ and $24$ years old) corresponding to the enrolment and calibration subsets of the original dataset.
The different paradigms for both datasets are presented in \autoref{fig:multidataset}. Other details about these datasets are also listed in \autoref{tab:datasets}.

\begin{table}[!ht]
\centering
\caption{Dataset, Cognitive task, time-locking event, time window and number of classes for all the sub-datasets in \erpcore and \mcv.}
\begin{tabular}{c|ccccc}
\hline
\textbf{Dataset} & \textbf{Cognitive Task ($\dataset$)} & \textbf{Time-Locking Event} & \textbf{Window size (ms) $(m)$} & \textbf{\# classes ($\Y$)} \\ \hline\hline
\multicolumn{1}{c|}{\erpcore} & N170 & Stimulus-locked & -200 to 800 & 2 \\
\multicolumn{1}{c|}{\erpcore} & MMN & Stimulus-locked & -200 to 800 & 2 \\
\multicolumn{1}{c|}{\erpcore} & N2pc & Stimulus-locked & -200 to 800 & 2 \\
\multicolumn{1}{c|}{\erpcore} & N400 & Stimulus-locked & -200 to 800 & 2 \\
\multicolumn{1}{c|}{\erpcore} & P3 & Stimulus-locked & -200 to 800 & 2 \\
\multicolumn{1}{c|}{\erpcore} & LRP & Response-locked & -800 to 200 & 2 \\
\multicolumn{1}{c|}{\erpcore} & ERN & Response-locked & -600 to 400 & 2 \\
\multicolumn{1}{c|}{\mcv} & ME & Stimulus-locked & 0 to 1000 & 3 \\
\multicolumn{1}{c|}{\mcv} & RS & Stimulus-locked & 0 to 1000 & 2 \\
\multicolumn{1}{c|}{\mcv} & SSS & Stimulus-locked & 0 to 1000 & 3 \\
\multicolumn{1}{c|}{\mcv} & TSS & Stimulus-locked & 0 to 1000 & 2 \\ \hline
\end{tabular}

\label{tab:datasets}
\end{table}

\subsection{Defining the decoding labels}

The two EEG datasets under investigation offer many options to define the
decoding
labels
$\y$.
For instance, one could consider the subject's response accuracy to the stimulus or whether the answer was in the appropriate time frame. 
Another possible label definition involves predicting the stimulus presented based on the recorded signal or the presence or absence of the task component \cite{VariationsERPCore:2023, kappenman2021erp}.
We describe hereafter how we defined the decoding labels in our experiments, starting with the \erpcore dataset.

The \emph{Flankers} task aims to elicit two possible ERP components: \emph{ERN} and \emph{LRN}. ERN
% is usually elicited by the tasks aiming to identify when
characterizes situations in which the subject makes an error, even when they are not consciously aware of them \cite{gehring2012error}.
% To do this, subjects must 
It is elicited here by asking subjects to
indicate whether they saw a left or right trigger. 
We assigned one label to trials where the subject's response matched the target stimulus, and another label when it did not.
If the trigger and response were both left or right, the trial was labelled as correct; otherwise, it was incorrect \cite{luck_2014, kappenman2021erp, luck_2022}.

The same Flankers task is used to elicit the
% Moving on to the 
\emph{LRN component},
% the evoked difference
which
is related to the side of the response hand. Unlike ERN, this component is associated with preparing the response rather than its execution, as reported in previous studies \citep{luck_2014, luck_2022}. Hence, we relied solely on the subject's response to determine the label for each trial in this case.
% the LRN component.
Specifically, if the subject answered with the left hand, we assigned one label; if they responded with the right hand, we assigned a different label.

% The usually elicited component in t
The \emph{Word Pair Judgment}
task is used to elicit the \emph{N400} component.
It consists in presenting participants with a prime word, followed by a target word and asking them to judge whether they are semantically related or not \cite{kutas2011thirty}.
To define the label for each trial, we use the stimulus definition. Specifically, if the presented words were semantically related, we assigned one label; if they were unrelated, we assigned a different label.

The \emph{N2pc} component is elicited through a \emph{Simple Visual Search} task
% The component in the \emph{Simple Visual Search} is the \emph{N2pc}. The cognitive task usually involves a neural component closely linked to a visual search task
\cite{luck2012electrophysiological}.
In this task, participants had to visually identify a specific target item (in blue or pink) among multiple distractors (in black).
Notably, each item is an outlined square with a gap on top or bottom, and subjects had to indicate the position of the gap on the target item (\cf \autoref{fig:multidataset}).
Despite the task assigned to the subjects, what is decoded in this experiment is the position of the target within the screen, \ie whether it appeared on the left or right side, since this is what the N2pc component really captures.
As usually done for this experiment~\cite{VariationsERPCore:2023}, we excluded trials with incorrect responses to ensure that the subjects were paying attention to the target.

% The \emph{Face Perception} usually contains the
The \emph{N170 component}~\cite{rossion2011erp}
% N170
was elicited by presenting stimuli in the form of cars, faces, and deformed versions of these objects, in what is called the \emph{Face Perception} task. 
Participants had to identify whether the presented stimulus was intact or scrambled. To simplify the analysis, we 
excluded scrambled stimuli
as recommended by the original authors of \erpcore~\cite{VariationsERPCore:2023}, 
and labelled the remaining ones according to whether they represented a car or a face.

Concerning the \emph{Active Visual Oddball} task,
% we usually have the \emph{P3b component} elicited \cite{p3b1969}. The usual cognitive task involved 
participants were asked to watch a sequence of random letters among A, B, C, D or E.
The first letter in a trial had the role of target, and subjects had to answer whether the following letters matched the target or not, which should elicit the \emph{P3b component}~\cite{p3b1969}.
To assign trial labels for this task, we considered two factors.
First, we checked whether the
% stimuli matched, \ie whether the
letter on the screen matched the target letter.
Second, we verified whether the participant provided the correct answer or not.
If both conditions were satisfied, we assigned one label to the trial. If not, we assigned another label.

Finally, the \emph{Passive Auditory Oddball} task aims to elicit the \emph{MMN component}~\cite{naatanen2004mismatch}. In our study, the stimuli were used to define the trials labels.
Specifically, we assigned one label if the individual heard a standard tone at 80 dB and another label if they heard a deviant tone
at 70 dB.

In the \mcv dataset, the definition of labels was more straightforward.
For instance, decoding labels for the \emph{RS} task simply corresponded to whether the subjects had their eyes open or closed.
For the \emph{TSS} and \emph{SSS} tasks, labels were directly defined based on the stimulus presented to the subjects, \ie whether it was a visual, auditory or somatosensory stimulation (3-class classification problem).
Likewise, decoding labels of the \emph{ME} task corresponded to the movement being executed by the subjects, \ie either the right foot, the right hand or the left hand (3-class classification problem).

\subsection{EEG pre-processing and epoching}

Both datasets were pre-processed 
following the authors' recommendations in all our experiments.
Namely, \erpcore recordings were filtered between $[0.5-40]$\,Hz with overlap-add FIR filtering.
Electric potentials were referenced on the average of electrodes P9 and P10 for all tasks except N170, for which we used the average of all 33 electrodes as commonly done in the literature~\cite{kappenman2021erp}.
Stimuli events were shifted $26$\,ms
forward in time
to account for the LCD monitor delay in the MMN task.
We downsampled the data from $1000$\,Hz to $250$\,Hz and used ICA~\cite{ablin2018faster}
to correct artefacts and discard particularly bad trials.
Finally, trials were cropped into $1000$\,ms windows based on the stimulus or response depending on the task, following values reported by \cite{kappenman2021erp}.
This dataset was entirely pre-processed using the \textsc{MNE-Python} library~\cite{mne:2013}.

For the \mcv dataset, we used the dataset pre-processed by the authors as described below.
Signals were band-pass filtered between 
$[0.01-200]$\ Hz
using a 4th order Butterworth filter and notch filtered between 
$[49-51]$\ Hz.
Potentials were referenced on the average of electrodes TP9 and TP10.
Visual inspection and ICA were used to remove artefacts, and bad channels were replaced by the average of the three neighbouring channels.
Finally, $1000$\,ms signals were cropped from each trial.
More pre-processing details can be found in \cite{Huang2022}.

\subsection{Data splitting}

The datasets were randomly split into a training, a validation, and a test set with respective proportions of $56\%$, $24\%$ and $20\%$.
Each split contains data from different subjects since we want to assess the cross-subject generalization of our models.
Trainings and evaluations were repeated with different splits following a 5-fold cross-validation scheme.

In the standard decoding experiments (\autoref{sec:dec-results}), models were trained and evaluated using data from the same cognitive task (\cf \autoref{fig:overview} C).
In the transfer learning experiments (\autoref{sec:transf-res}),
models were pre-trained on training subjects of some task $A$, then fine-tuned through linear probing on data from the same subjects performing a different task $B$ and finally evaluated on unseen test subjects carrying out this same task $B$, as described in \autoref{sec:linear-probing} and illustrated in \autoref{fig:overview} D.
Thanks to this data splitting alignment across cognitive tasks, we ensure that the test subjects remain the same after transfer,
avoiding any leakage between test and training sets.

\subsection{Decoding models}

%%%%%%%%%%%%%%%

We evaluated the transferability of learned tasks using three state-of-the-art deep learning models: ShallowNet~\cite{braindecode:2017}, EEGNet~\cite{lawhern2018eegnet}, and EEGInception~\cite{santamaria2020eeg}. ShallowNet is an efficient network inspired by the Filter Bank Common Spatial Pattern method \cite{filterbank:2008}.
It contains $36k$ trainable parameters. EEGNet model has a convolutional layer for channel-based EEG data filtering, a depth-wise convolutional layer acting as a spatial filter across channels, a separable convolutional layer for categorical feature extraction, and a fully connected layer as classification head. Finally, EEGInception has $15k$ parameters and extracts low- and high-frequency features in parallel at different scales.
These architectures delivered state-of-the-art results in several EEG decoding tasks and datasets~\cite{tsiouris2018long, borra2020interpretable, zancanaro2021cnn, cai2021deep, bakas2022team, wei2021beetl}.
In all our experiments, we used the implementations of these models available in the \textsc{Braindecode} library~\cite{braindecode:2017} with default hyperparameter values.

As baselines, we also tested Riemannian methods, which are known as the best machine learning techniques for BCI and ERP, together with DL models.
To this end, we evaluated both Minimum Distance to Riemannian Mean (MDM)~\cite{chevallier2018brain, barachant2010riemannian} and Tangent Space methods with either logistic regression or support vector machine classifiers.
We also evaluated these methods with a varied set of covariance matrices, such as ERPCovariances~\cite{barachant2014plug} and Xdawn~\cite{khazem2021minimizing}.
These methods were implemented and trained using the \textsc{MOABB} and \textsc{PyRiemann} library~\cite{moabb2018, pyriemann}.

\subsection{Training setting}

All DL models were initialized with the Xavier sampling~\cite{glorot2010understanding} and trained with the AdamW optimiser~\cite{loshchilov2018decoupled} using default parameters $\beta_{1}=0.9$ and $\beta_{2}=0.999$
and a weight decay of $5 \times 10^{-4}$.
The initial learning rate was set to $10^{-4}$
for ShallowNet and EEGInception, and to $6.25 \times 10^{-4}$ for EEGNet.
Trainings lasted at most 200 epochs and an early stopping~\cite{prechelt1998early} procedure with a patience of 50 epochs was used. The training of all deep learning models was carried out using \textsc{Pytorch} and \textsc{Braindecode} libraries ~\cite{pytorch:2019, braindecode:2017} on an Nvidia DGX with 4 A100 boards.
%The source code and trained models are publicly available at \url{https://github.com/anonymous}\footnote{The code will be made available after acceptance of the manuscript}.

\section{Results and Findings}\label{sec:resu}

\subsection{Decoding perfomance}\label{sec:dec-results}

\begin{figure}[!ht]
    \centering
  \subfigure[\erpcore dataset]{
    \includegraphics[width=\linewidth]{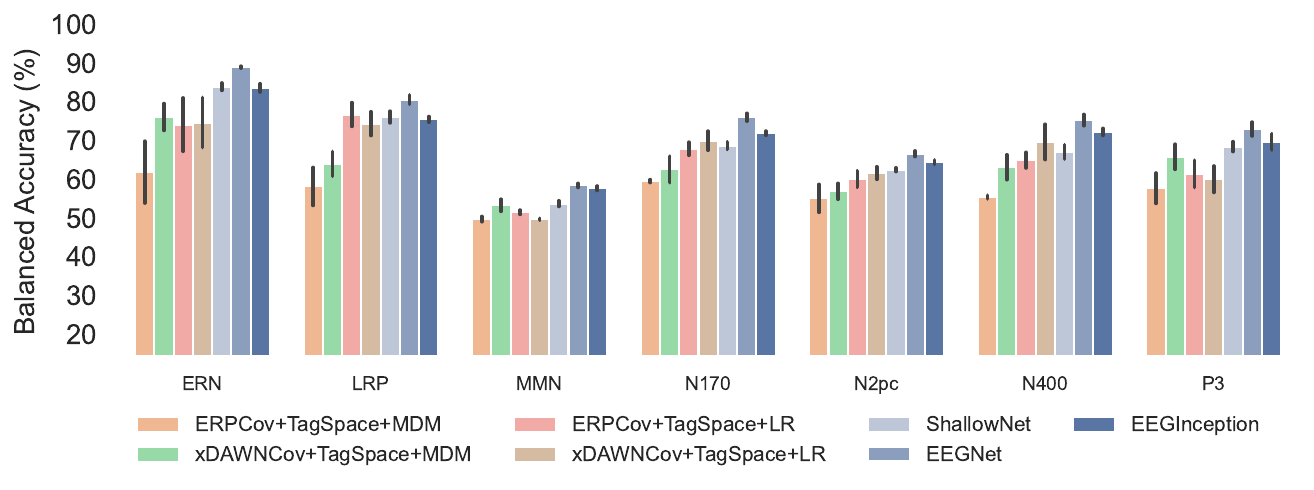}
    \label{fig:erpcore_deep_baseline}
  }
  \subfigure[\mcv dataset]{
    \includegraphics[width=\linewidth]{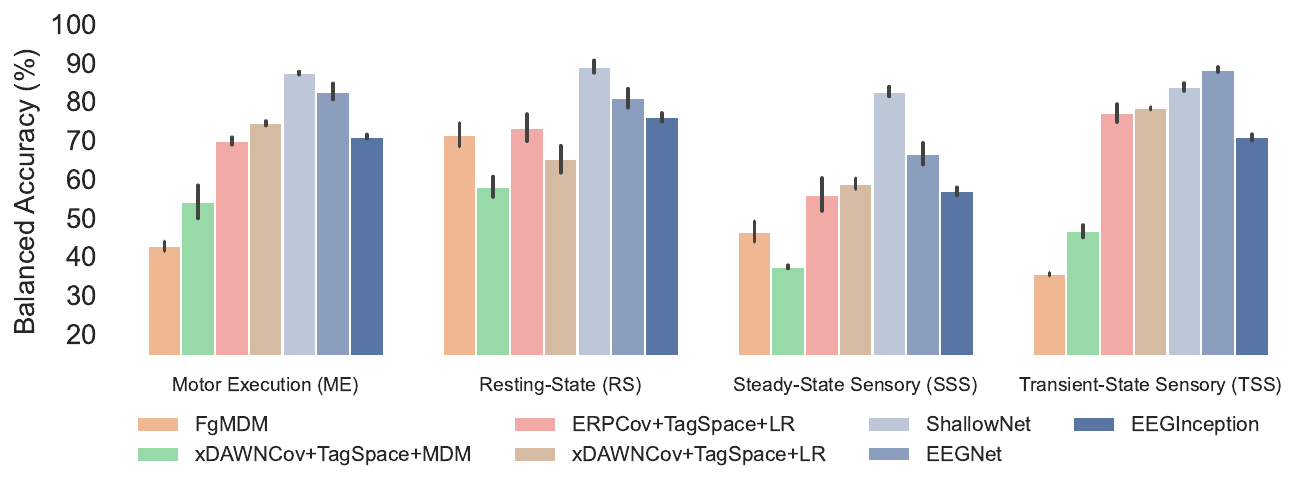}
    \label{fig:m3cv_deep_baseline}
  }
    \caption{
    Cross-subject balanced accuracy across paradigms.
    Error bars correspond to the standard-deviation across 5-fold cross-validation.
    DL methods outperform machine learning baselines on most paradigms.
    EEGNet is consistently better than other architectures in \erpcore.
    ShallowNet leads to the best scores across \mcv paradigms, except for TSS.
    }
    \label{fig:deep_baseline}
\end{figure}

We first analyse the decoding performance of studied models on \erpcore and \mcv datasets in a standard decoding setting, without any transfer.
Overall, DL methods deliver superior balanced accuracies than baseline methods, as shown in 
\autoref{fig:deep_baseline}.

Their scores also exhibit lower standard deviations than machine learning methods, demonstrating the stability of learned representations across subject splits.
Notably, EEGNet consistently led to the best performance on \erpcore for all paradigms, whereas ShallowNet outperformed others on three out of four paradigms of the \mcv dataset.
Also, note that our results on \erpcore are consistent with those reported in \citep{bonasch2023decoding, trammel2023decoding}.

\begin{figure}[t]
  \centering
  \subfigure[\erpcore dataset]{
    \includegraphics[width=0.45\linewidth]{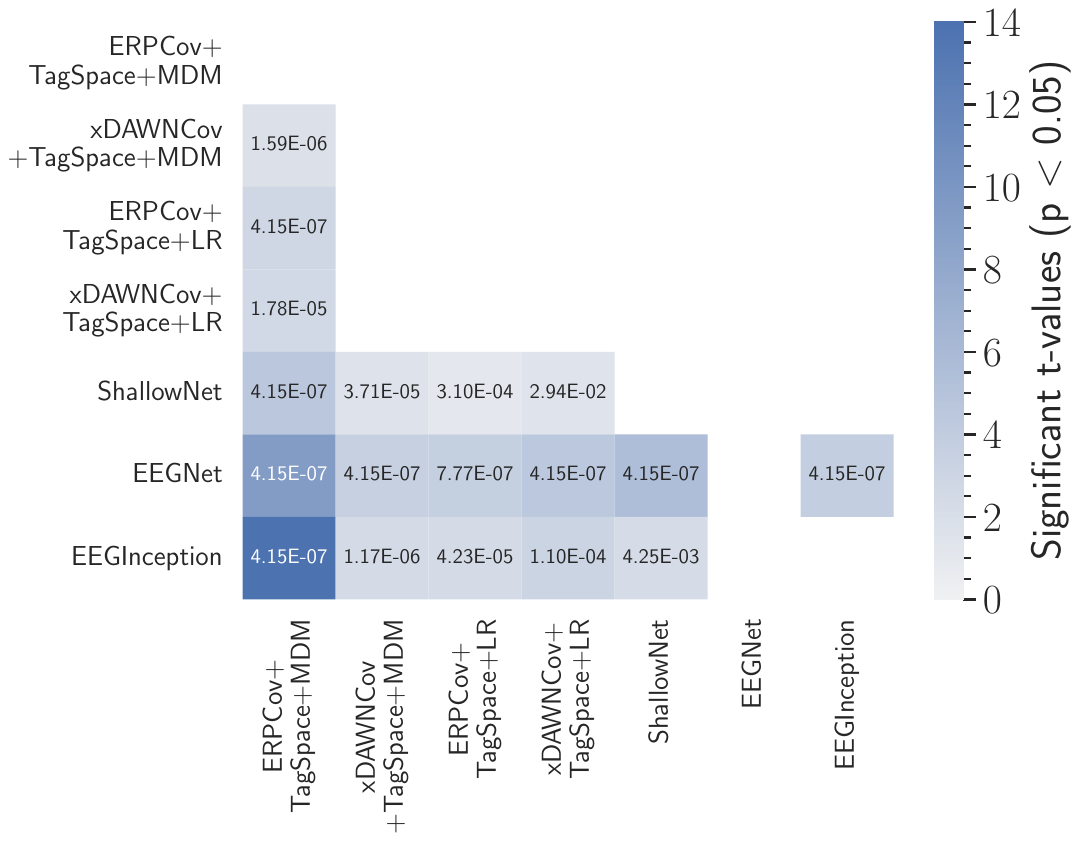}
    \label{fig:stats_erpcore}
  }
  \subfigure[\mcv dataset]{
    \includegraphics[width=0.45\linewidth]{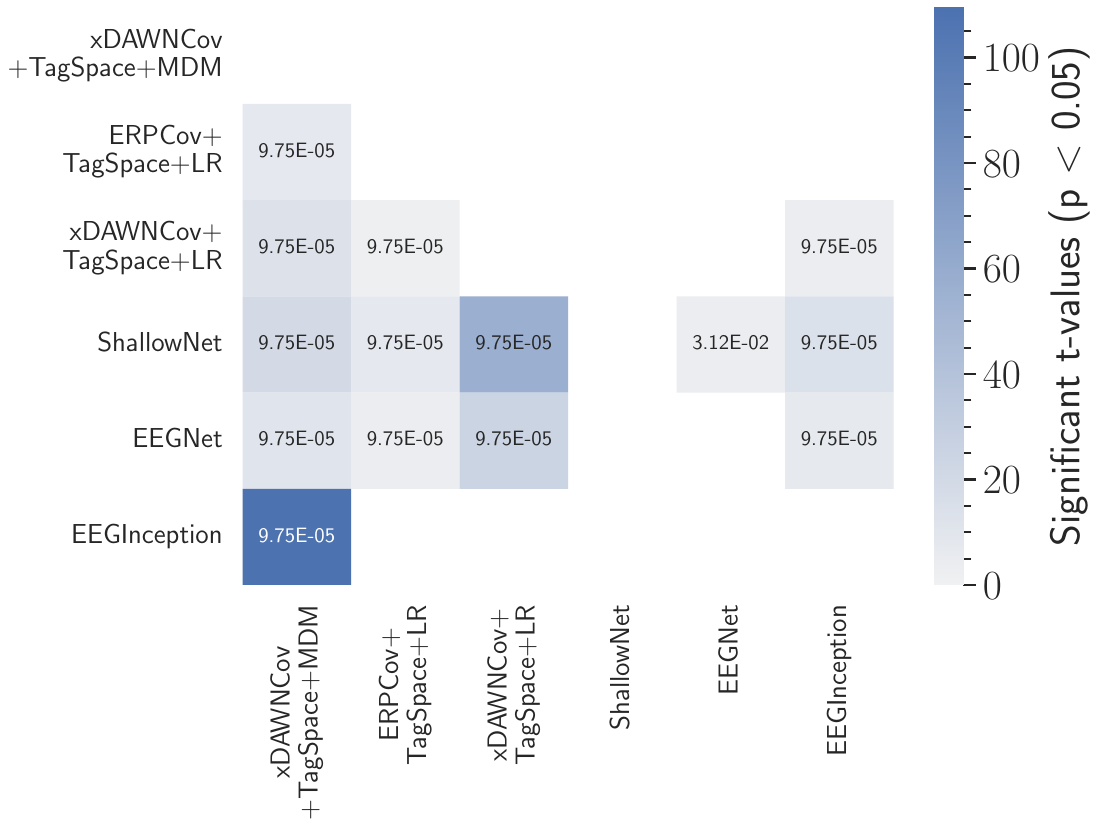}
    \label{fig:stats_m3cv}
  }
  \caption{
  Significant pairwise standardized mean performance difference between decoding models from \autoref{fig:deep_baseline}.
  Statistical significance was computed using a corrected permutation test, with a cutoff p-value of $0.05$.
  }
  \label{fig:stats_both}
\end{figure}

To support this analysis,
we performed a permutation signed-rank test for each model pair within each paradigm to determine
whether observed performance gaps are significant (\autoref{fig:stats_both}).
The resulting p-values were combined using Stouffer's method, with a weighting given by the square root of the number of subjects,
and a Bonferroni correction was applied to account for multiple comparisons, as done in \cite{moabb2018}. The standardized mean difference was calculated within each dataset to determine the effect size. Examining Figure \ref{fig:m3cv_deep_baseline}, we observe that EEGNet slightly outperforms ShallowNet, although the difference is not statistically significant, as highlighted in Figure \ref{fig:stats_m3cv}.

From a paradigms perspective, the most challenging task within the \erpcore dataset is the MMN paradigm, which is unique in that the subject does not have an active response moment.
Conversely, the least difficult tasks correspond to the ERN and LRP.

Overall, the best DL architectures (EEGNet for \erpcore and ShallowNet for \mcv) lead to similar performances across paradigms, even in scenarios with different numbers of classes (\eg two classes in RS vs. three classes in ME, SSS and TSS).
This suggests that the results are robust and confirms that the selected decoding architectures are well-suited to the different experimental paradigms studied in this work.

\subsection{Transfer learning performance} \label{sec:transf-res}

\subsubsection{\erpcore}

\begin{figure}[!ht]
    \centering
   \includegraphics[width=\linewidth]{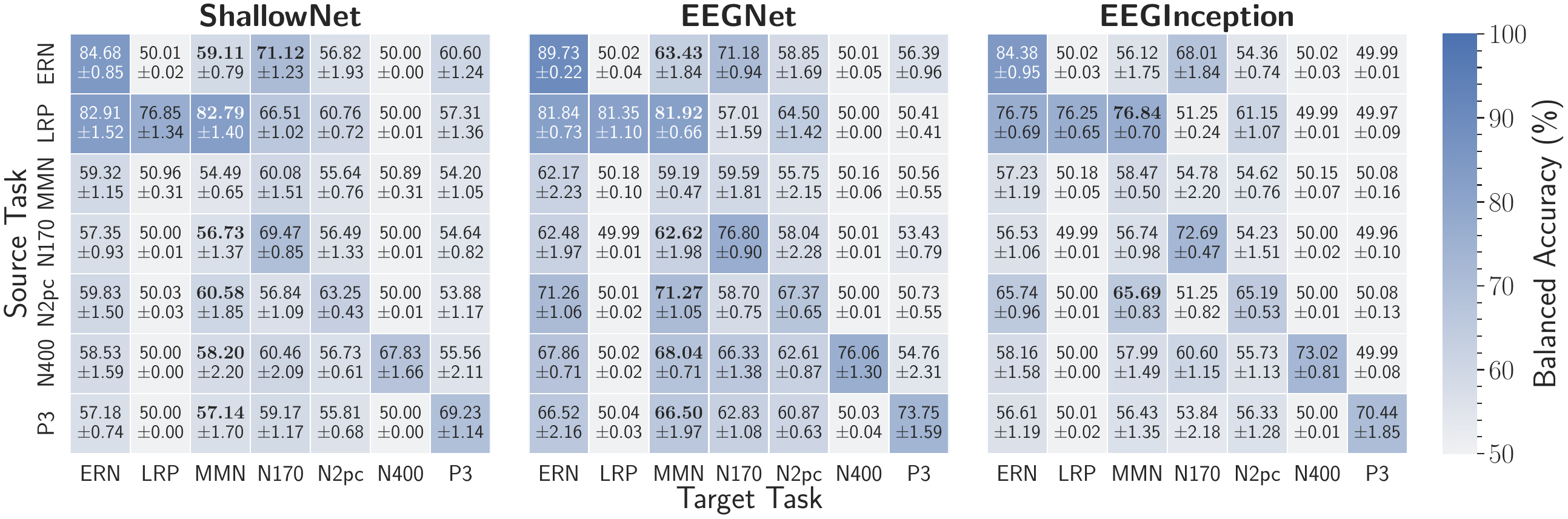}
    \caption{
    Transfer balanced accuracy for each pair of sources and target cognitive tasks
    in the \erpcore dataset.
    Cell values correspond to average performance and standard-deviation across 5-fold cross-validation.
    Diagonal values correspond to standard decoding balanced accuracies, without transfer (\cf \autoref{sec:dec-results}).
    }
    \label{fig:erpcore_confusion}
\end{figure}

We now analyse the transfer performance between paradigms to quantify how transferable each cognitive task is in relation to one another.
\autoref{fig:erpcore_confusion} shows the obtained balanced accuracy for each pair of source and target cognitive tasks in the \erpcore dataset.
A first intriguing
observation
is the asymmetry
of the transfer matrices obtained,
regardless of the model used,
indicating that the transferability between tasks is highly directional.
We can see that some tasks do not transfer well, leading to accuracies close to chance ($50\%$ in this dataset, as all classification problems are binary).
Notably, no source task is correctly transferring to either to LRP or N400 paradigms.
This means that these cognitive tasks rely on very specific representations that are not 
elicited by the other
tasks and are hence erased by models pre-trained on other paradigms.

Conversely, representations learned on some source tasks seem useful for decoding others, at least to a level of accuracy comparable to a model fully trained and evaluated in a standard fashion (diagonal values in the figures). 
For example, LRP proves to be a good source task
when transferring to ERN and MMN.
This suggests that LRP elicits some brain activations common to ERN and MMN, which are learned by the decoding networks into its hidden representations.

More strikingly, pre-training on some source tasks, such as ERN, improves performance on some other target tasks, like N170, even \emph{exceeding} the reference performance of a model fully trained on this target task.
The best example of this is the MMN task, whose performance is boosted beyond the reference accuracy when transferring \emph{from all possible source tasks}.
With ShallowNet,
for example,
we observe improvements of $28.3\%, 6.1\%, 4.6\%, 3.7\%, 2.7\%,$ and $2.2\%$ for the LRP, N2pc, ERN, N400, P3, and ERN source tasks, respectively.
This is also observed for EEGNet and EEGInception, although more discretely.

In order to better visualize these complex connections between tasks, we processed the matrices from \autoref{fig:erpcore_confusion} into transferability maps, shown in \autoref{fig:network_graph_erpcore}.
Arrows' widths are proportional to corresponding transfer accuracies, where performances close to chance level were omitted.
More precisely, they correspond to transferability scores $s_{S,T}$ for each source and target tasks, obtained by linearly rescaling the corresponding accuracies $a_{S,T}$ so that $0$ corresponds to the chance level $c_T$ and $1$ corresponds to the reference accuracy without transfer (\ie matrix diagonal $a_{T,T}$):
\begin{equation} \label{eq:accs_rescale}
    s_{S,T} = \frac{\max (a_{S,T} - c_T, 0 )}{a_{T,T} - c_T}\,.
\end{equation}
The Scalable Force-Directed Placement (SFD) layout algorithm~\cite{hu2005efficient} was used to create the graphs.
This figure clearly shows that LRP is a very good source task and that MMN is a great target task.
Note that all three different architectures lead to consistent maps overall, evidencing 
the stability of our findings.

\begin{figure}[htp]
\centering
\subfigure[\erpcore]{
\includegraphics[width=\linewidth]{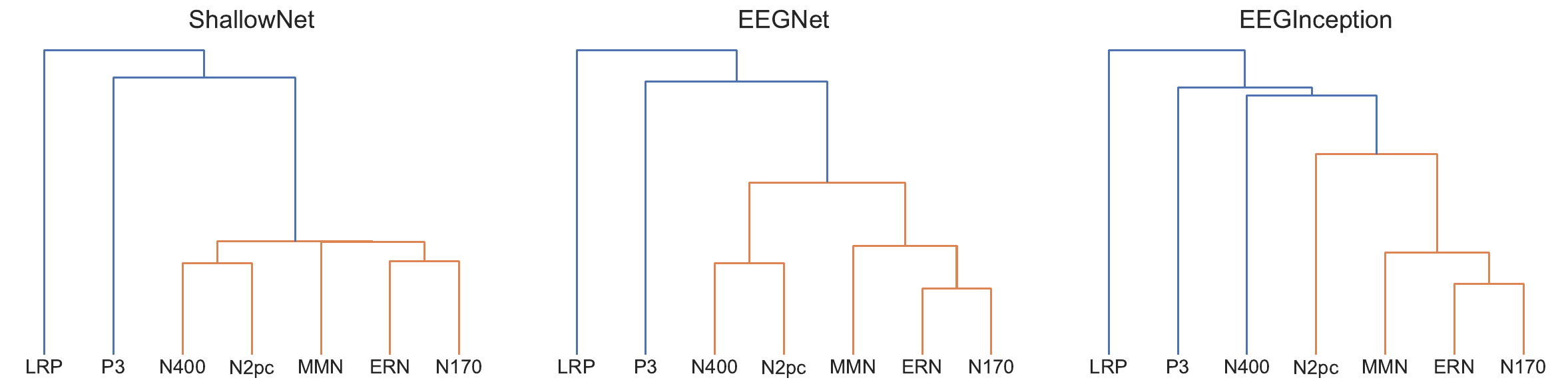}
\label{fig:dendogram_erpcore}
}
\subfigure[\mcv]{
\includegraphics[width=\linewidth]{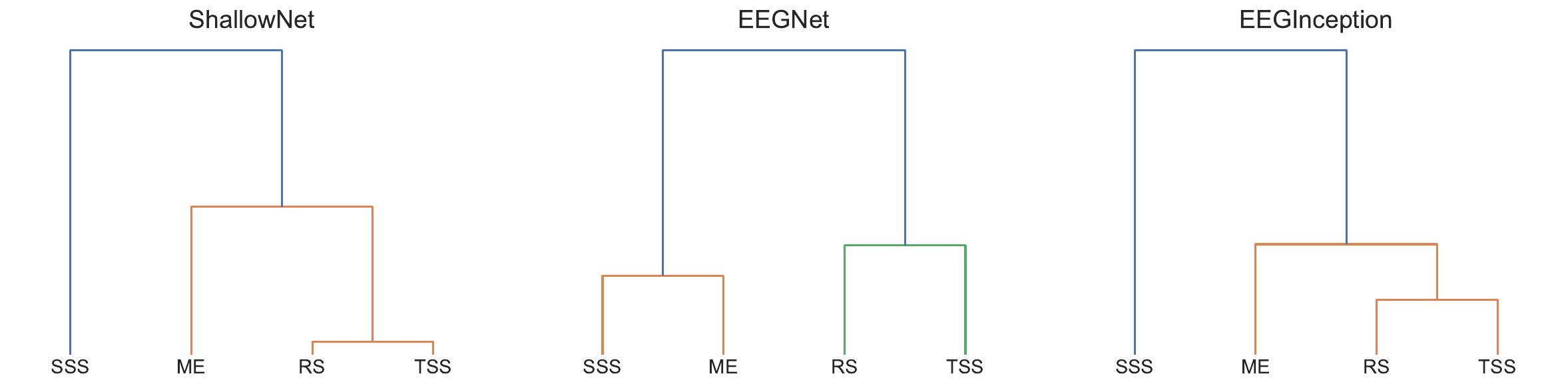}
\label{fig:dendogram_m3cv}
}
\caption{
Hierarchical clustering dendrogram, showing the resemblance between source tasks in both datasets.
Each task is represented by its row in the transferability matrices (Figures \ref{fig:erpcore_confusion} and \ref{fig:m3cv_confusion}) rescaled through \autoref{eq:accs_rescale}.
Pairwise distances were than computed using the Euclidean distance,
before clustering the tasks with the 
UPGMA algorithm \cite{Sokal1958ASM}.
}
\label{fig:dendogram_erpcore_m3cv}
\end{figure}

As another way to compare tasks, we also clustered them as done in \cite{zamir2018taskonomy} by representing each task by its row vector in the transferability matrix rescaled through \eqref{eq:accs_rescale}.
This led to the dendrograms depicted in Figure \autoref{fig:dendogram_erpcore}.
By comparing tasks through their row vectors, we are trying to see whether they transfer in a similar fashion to other tasks.
Most tasks cluster together, with N400-N2pc and N170-ERN being the closest pairs.
We also see that LRP stands outside of the cluster, as it is a particularly good source task.

\subsubsection{\mcv}

\begin{figure}[ht]
    \centering
    \includegraphics[width=\linewidth]{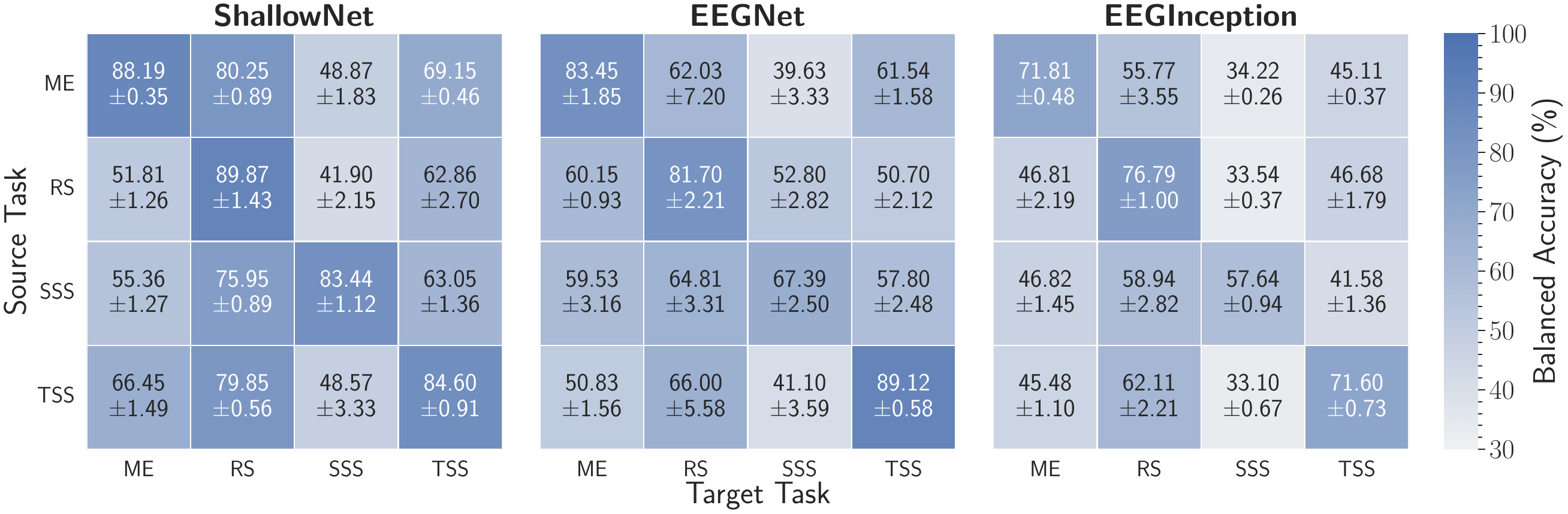}
    \caption{Average performance of the three deep learning models for the transfer learning setting in the \mcv Dataset.}
    \label{fig:m3cv_confusion}
\end{figure}

Moving on to \mcv, we see in \autoref{fig:m3cv_confusion} that we also obtain very asymmetrical transferability matrices for this dataset.
The best example of this asymmetry are the tightly related SSS and TSS tasks, which share the same class labels: visual, auditory or somatosensory stimuli.
Surprisingly, while SSS transfers relatively well to TSS, the opposite is not true, showing that steady-state representations have something specific which is not captured when pre-training on transient state data.

Another striking observation is that the EEGInception matrix differs from ShallowNet and EEGNet ones.
This is probably due to the lower pre-training performance of this model seen on Figure \autoref{fig:m3cv_deep_baseline}, evidencing that it learned weaker representations.
Interestingly, we see on both \erpcore and \mcv results that transfer performances across models (Figures \ref{fig:erpcore_confusion} and \ref{fig:m3cv_confusion}) rank consistently with pre-training performances (\autoref{fig:deep_baseline}).

From a paradigms perspective, we can see that no transfer accuracies exceed the reference diagonal values in this dataset.
Nonetheless, RS appears as a great target task, benefiting from representations learned in all source tasks.
To a smaller extent, TSS also exibits good performances for all source tasks when using the ShallowNet architecture.
We hypothesize that this is only visible for this architecture because it is the only one capable of extracting the information shared by all 4 paradigms, given its superior decoding performance for this dataset (\cf \autoref{fig:m3cv_deep_baseline}).
On the contrary, it appears that it is very difficult to transfer onto the SSS paradigm, the extreme case being obtained with EEGInception, for which the chance accuracy is attained for all sources ($33\%$ in this case, since it is a 3-class classification problem).

\begin{figure}[t]
  \centering
  \subfigure[ShallowNet]{
    \includegraphics[width=0.3\linewidth]{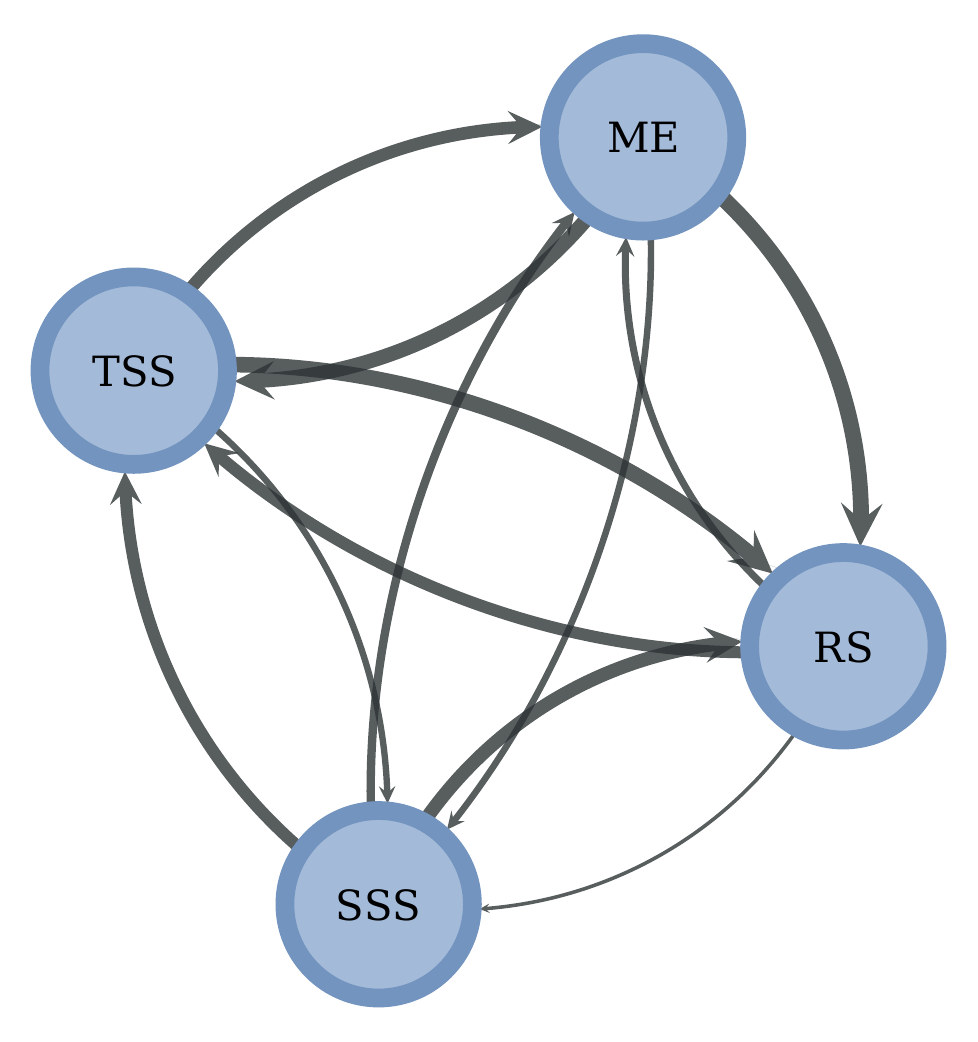}
  }
  \subfigure[EEGNet]{
    \includegraphics[width=0.3\linewidth]{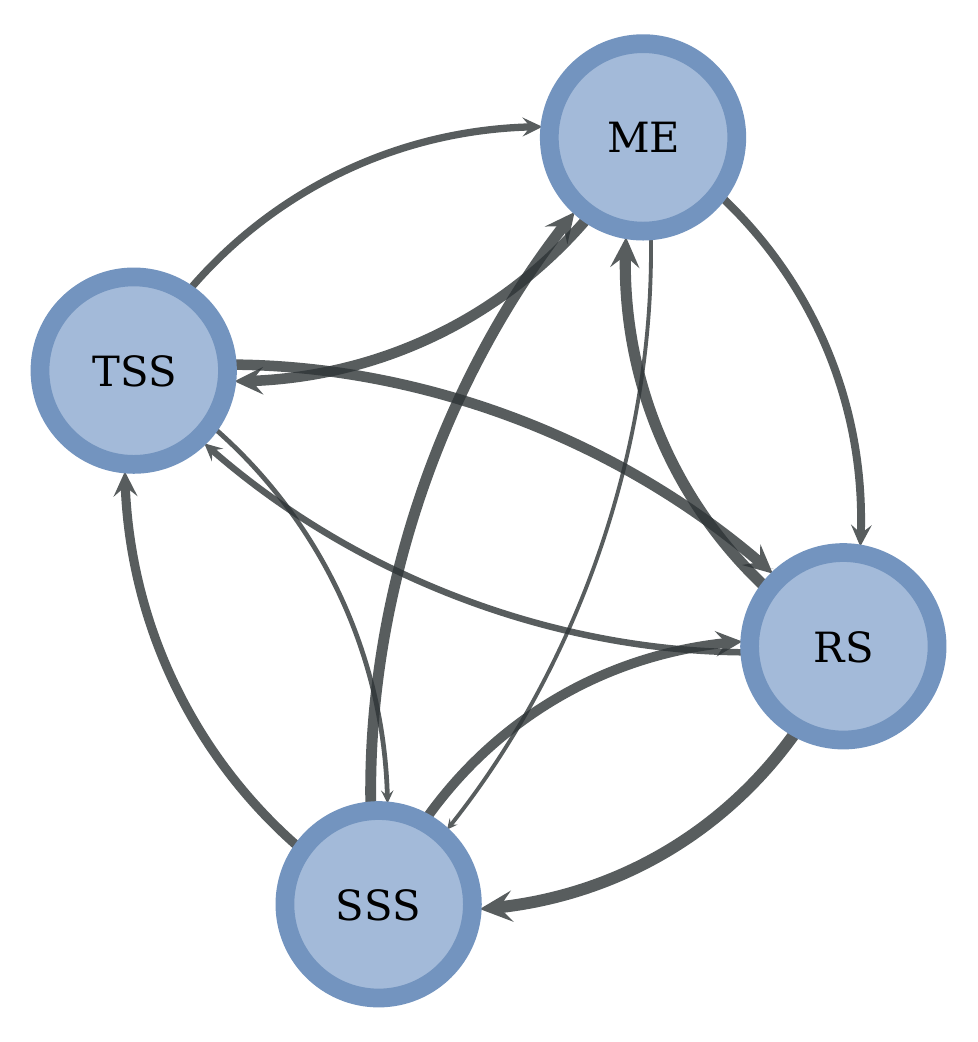}
  }
  \subfigure[EEGInception]{
    \includegraphics[width=0.3\linewidth]{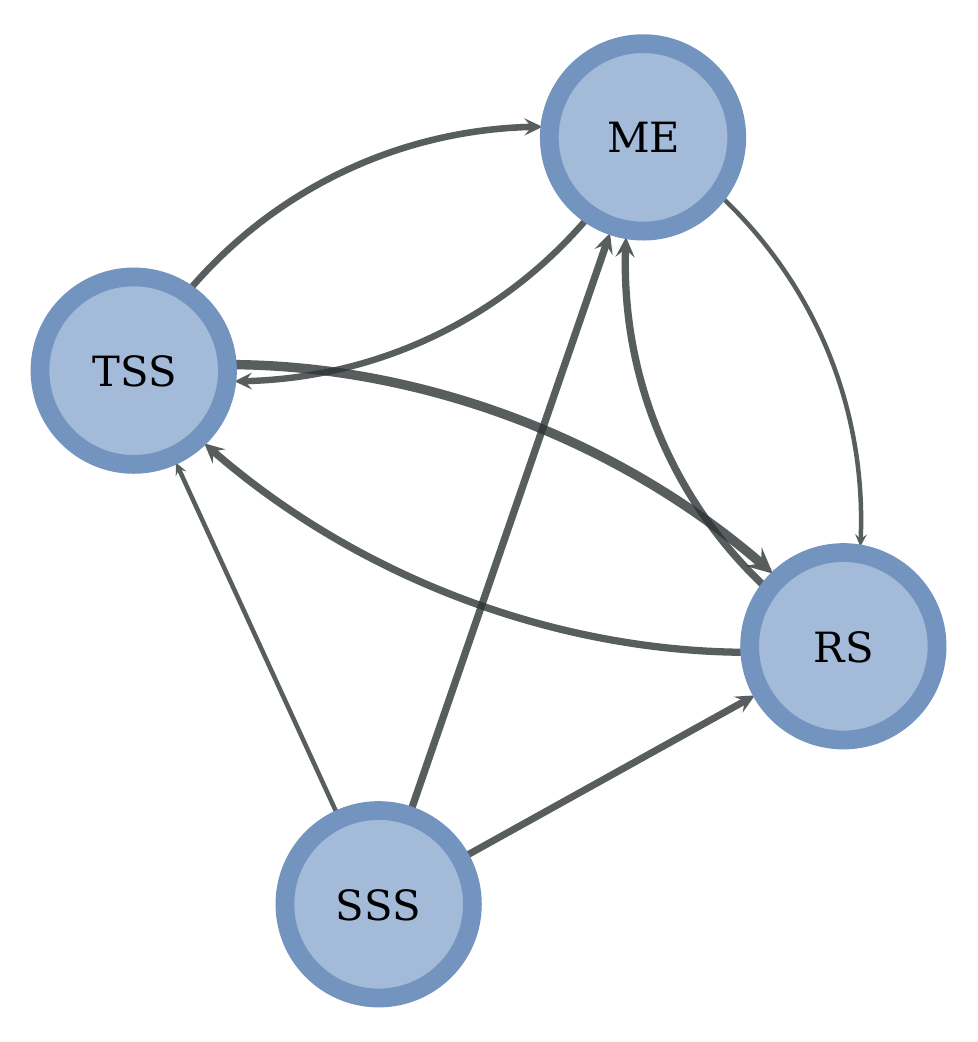}
  }
  \caption{
  Transferability map of the \mcv dataset.
  Each node corresponds to a distinct paradigm, while arrows
  width represent the average transfer performance when using the representations learned from a source task to decode a target task.
  All tasks transfer well to RS.
  While SSS transfers to TSS, the opposite is not true.
  Surprisingly, ME and TSS transfer well to each other in both ways.
  }
  \label{fig:network_transf_m3cv}
\end{figure}

As for \erpcore, we computed the transferability maps for \mcv in \autoref{fig:network_transf_m3cv}.
This visualization allows to clearly see the
consistent connections
from TSS, ME and SSS towards RS, regardless of the model employed.
The same pattern is visible for TSS with ShallowNet, which receives broad arrows from all neighbouring tasks.
We also note a persistent bidirectional connection between TSS and ME, which is quite surprising given that they are very different tasks.

The similarity between the best target tasks, RS and TSS, becomes even more apparent on the 
clustering dendogram from Figure \autoref{fig:dendogram_m3cv}
Indeed, these paradigms are grouped in their own cluster with EEGNet and appear as the closest paradigms with ShallowNet and EEGInception. 
This figure also confirms that SSS is a very particular task, as it is outside the main cluster for two out of three dendograms.

\section{Discussion and impact}
\label{sec:discussion}

Our analyses have yielded a reliable cognitive map through the use of transfer learning and its successful generalization across various cognitive paradigms. The resulting taxonomy maps, which encapsulate the knowledge of model parameters, open up diverse possibilities for practical applications and further research. In this discussion, we will delve into some potential applications of cognitive mapping from the perspective of machine learning models, emphasizing the benefits and implications they offer.

First and foremost, these cognitive maps 
% (\emph{domain})
can significantly enhance and optimize the training of predictive models. As exemplified in this study, one compelling approach is to leverage a task that exhibits strong transferability as a template for decoding challenging cognitive tasks, as shown for MMN or N170 ERPs. This aspect proves especially crucial in psychophysical experiments, where subject numbers are limited while experimental variables are abundant. These factors seem to be observed in other areas, such as computer vision \cite{InfluenceCrossDataset2022}.

Additionally, the cognitive taxonomy maps can play a pivotal role in the design of improved Human Machine Interfaces (HMIs) for Brain-Computer Interface (BCI) applications. Our focus particularly centers around integrated BCIs~\cite{lotte2018review}, which extend beyond mere control functions and delve into scenarios involving error detection and negative potential. By incorporating insights gained from cognitive mapping, we can significantly enhance the performance and effectiveness of HMIs in such contexts. For example, the detection of ERN could help to mitigate situations where the BCI system selects unwanted commands, resulting in user frustration.

Furthermore, the clinical applications of cognitive mapping hold immense potential, especially in aiding patients with schizophrenia who encounter difficulties in emotion detection and facial recognition \citep{Watson:2013}. Leveraging the insights derived from the N170 task \citep{Yang:2020}, we can devise interventions that boost performance and facilitate improved outcomes for individuals grappling with these challenges. This is especially useful in the context of remediation of emotion recognition, where schizophrenic patients have difficulties to recognise faces and to extract relevant cues (like eyebrows or mouth movements) for interpreting social interaction adequately. Objective information indicating that the patient has recognised a face during an interaction, instead of focusing on irrelevant details for instance, could help therapists for rehabilitation purposes.

Moreover, our findings shed light on the functional networks underlying related tasks. Notably, closely related tasks often engage similar functional networks, suggesting the presence of shared evoked components~\cite{luck_2014,luck_2022}. This implies that activations observed in one task can enhance and refine the performance of another task. While these tasks may exhibit distinct temporal dynamics and spatial orientations, they ultimately serve the same overarching purpose and engage overlapping cortical networks. Unraveling the sources responsible for generating event-related potentials (ERPs) becomes more feasible when considering these perspective-dependent variations in activation patterns. This understanding illuminates the intricate interplay among different components of the cortical network.

One compelling avenue for applying the insights gained from our study lies in the realm of source localization, a classic inverse problem in neuroscience~\cite{mne:2013}. Cognitive mapping can serve as a potent constraint in reconstructing brain activity, leveraging functional regularization. Source localization methods rely on physical and anatomical constraints to yield plausible solutions. Information regarding share evoked components could help to define functional constraints on estimated solution. By leveraging the connections uncovered through cognitive mapping, we can construct robust source localization models with enhanced accuracy and interpretability. 

Finally, we can address the issue of BCI illiteracy, which refers to the challenges faced by individuals to effectively operate BCIs~\cite{lee2019, SUN2022111}. Our transfer learning approach, which could be applied at the subject level, holds promise in mitigating this challenge. BCI illiteracy is mostly observed on few specific tasks for a given user while accuracy on other tasks are correct~\cite{lotte2018review}. 
Based on our results, a first possible approach is to apply task transfer, training on an effective decoding task for a user to transfer the results to an inefficient task. A second and more exploratory approach is to investigate user transfer, that is 
transfer the most effective decoding task from one subject and apply it to another with poor decoding abilities. This area presents an exciting and promising avenue for further exploration and improvement.

\section{Conclusion}

Our study investigates the transferability of deep learning representations
through extensive experiments on two EEG datasets. 
To our knowledge, our work is the first to construct representation transfer maps for EEG decoding.
These maps reveal a complex and asymmetric hierarchical relationship between cognitive tasks, enhancing our understanding of brain decoding and neural representations.
Our findings also have very practical implications for mitigating data scarcity, demonstrating performance improvements in real world data.

One limitation of our work is that we focus on ERP and BCI datasets.
Future work could hence extend our analysis to a broader range of cognitive tasks, or even investigate
% We could also imagine the investigation of
the transferabily across datasets that do not share the same cohort of subjects.
% About the future work and limitations, first, our study focused on ERP and BCI datasets, limiting the exploration of additional multi-cognitive task datasets. Expanding our research would require a larger dataset or data collection efforts. 
Other interesting avenues for investigation could be the study of other types of transferability,
for example by carrying out a similar analysis in a self-supervised setting.
% Substituting the final layer of the neural network with a more sophisticated counterpart could enable the evaluation of alternative adjustment methods beyond linear probing. Enhancing the decoding layer may lead to a more compelling and structured common representation.
% Additionally, exploring the implementation of a self-supervised framework could uncover new avenues for investigation.

\section*{Acknowledgments}

We would like to thank Dr André Cravo for his valuable feedback on this manuscript.
This work was supported by the ANR BrAIN (ANR-20-CHIA-0016), ANR AI-Cog grants (ANR-20-IADJ-0002), DATA IA (project YARN). The work of BA was supported in part by the CAPES under Grant 001 and Data IA mobilité internationale. The work of WHLP was supported by Wellcome Innovations under Grant [WT213038/Z/18/Z]. 

%Bibliography
\bibliographystyle{unsrtnat}  
\bibliography{templateArxiv}

\end{document}